\documentclass[12pt]{article}
\usepackage{amssymb,latexsym,amsmath,amsfonts,multicol}
\usepackage[margin=1in]{geometry}
\usepackage{graphicx}
\usepackage{float}
\usepackage{upgreek}
\date{}
\begin{document}
\title{Lie Symmetry Analysis and Some New Exact Solutions to the KP-BBM Equation}
\author{ Arindam Ghosh\footnote{mail:arindamghosh227@gmail.com}, Sarit Maitra\footnote{mail:sarit2010.nt@gmail.com}\\
\newline Dept. of Maths.\\ National Institute of Technology Durgapur, India. }     
\maketitle
\begin{abstract}
This paper is aimed to study the KP-BBM equation, which was proposed by Abdul Majid Wazwaz [Wazwaz. A.M.: Applied Mathematics and Computation, 169 (2005), 700–712.]. To check its integrability Painlev\'e test has been performed. Lie Symmetry analysis has been done and point symmetry generators are obtained. The invariants of the Lie algebra are found and the one dimensional optimal system for subalgebras of the obtained Lie algebra is constructed by using the Hu-Li-Chen algorithm. Three similarity reductions and corresponding exact solutions are derived. Also Homogeneous balance method, $Tanh$ method are used to find exact solutions. Solitary wave like solutions are obtained and plotted for some suitable values of the parameters involved. The effects of the nonlinear coefficient and dispersion coefficient on the obtained solitary waves are discussed.
\end{abstract}
\section{Introduction:}
Nonlinear differential equations (NDE) have become one of the most essential tools to study many problems of different branches of physics and mathematics \cite{lakshmanan2012nonlinear, strogatz1994nonlinear}. It is used to formulate different models of astrophysics, cosmology, hydrodynamics, plasma physics, nonlinear optics, mathematical biology \cite{strogatz1994nonlinear, jordan2007nonlinear}. Integrability provides significant information for physical phenomena described by NDEs \cite{goriely2001integrability}. A robust and profound tool for checking the integrability of a differential equation is the Painlev\'e test \cite{goriely2001integrability, conte2008painleve}. Though it was introduced by Painlev\'e and his followers on ordinary differential equations, the concept was extended in the context of PDE by Weiss, Tabor and Carnevale \cite{weiss1983painleve}. The importance of integrability, Painlev\'e property in the study of differential equations are well discussed in \cite{ramani1989painleve}. On the other hand exact solutions of NDEs provides a lot of quantitative information to these problems. In last few decades mathematicians proposed numerous techniques in this direction. Among them homogeneous balance method \cite{wang1996application, maitra2019exact}, Lie symmetry method \cite{hydon2000symmetry, cantwell2002introduction}, $Tanh$ method \cite{malfliet1996tanh}, sub ODE method \cite{zhouexact}, Hirota's bilinear method \cite{hirota2004direct}, First integral method \cite{ghosh2021first} are widely used for finding exact solutions. Homogeneous balance (HB) method is introduced by Mingling Wang \cite{wang1995solitary}. Wang et al. obtained exact solutions of some well known nonlinear partial differential equations by using HB method in \cite{wang1996application}. An extended version of this method is reported by Maitra et al. while obtaining a system of solutions of the SIDV equation \cite{maitra2019exact}. The $Tanh$ method, first introduced by Malfliet and Hereman \cite{malfliet1996tanh}, is widely applied to find $Sech^2$ solutions of NDEs. The First integral method, another useful method, first introduced by Zhaosheng Feng \cite{feng2002first} based on theory of commutative algebra. A vivid and rigorous description with application of this method is given in \cite{ghosh2021first}  by Ghosh et al.

  Lie group of symmetry method, developed by Sophus Lie, is very useful and one of the classic techniques of solving differential equations \cite{hydon2000symmetry}. Corresponding to each subgroup of the Lie group, the differential equation can be reduced to a differential equation with less independent variables which possesses group invariant solutions. As almost always there exists infinite number of such subgroups \cite{olver1993applications}, it is not possible to list all such group invariant solutions. For this reason the necessity of finding group invariant solutions that are inequivalent to each other came into account which leads to the concept of optimal system of subalgebras \cite{olver1993applications}. In application, to each member of the optimal system group invariant solutions are found. By using the invariants of the Lie algebra, Hu-Li-Chen proposed an effective algorithm for finding optimal system of subalgebras \cite{hu2015direct}. Beside finding exact solutions, conservation laws and many other important properties of differential equations can be constructed by using the knowledge of symmetry \cite{cantwell2002introduction}. 

In this work we are going to study the KP-BBM (Kadomtsev-Petviashvili-
Benjamin-Bona-Mahony) equation, in the context of Lie symmetry analysis, exact solutions, Painlev\'e test:
\begin{eqnarray}
& &(u_t+u_x+a(u^2)_x+bu_{xxt})_x+ku_{yy}=0\label{1} \\ 
\Rightarrow & & u_{xt}+u_{xx}+2a(u_x)^2+2auu_{xx}+bu_{xxxt}+ku_{yy}=0,\label{2} \\& &\nonumber
\end{eqnarray}
which was introduced by Abdul Majid Wazwaz \cite{wazwaz2005exact}, where $a, b$ and $k$ are the coefficients of nonlinear, dispersion and dissipation terms respectively. Wazwaz derived this equation from the BBM (Benjamin-Bona-Mahony) equation \cite{wazwaz2005exact}, which is used as a model for propagation of long waves. He found compactons, solitons and periodic solutions. Zhou et al. found some exact travelling wave solutions by generalized sub ODE method in \cite{zhouexact}. Using Hirota bilinear method Manafian et al. have obtained periodic wave solutions \cite{manafian2020periodic}. 
 
 The contents of this article is arranged as follows: in Section 2 and 3  Painlev\'e test and Lie symmetry analysis have been performed on the KP-BBM equation respectively. Exact solutions are obtained in Section 4 by homogeneous balance method, in Section 5 by $Tanh$ method.  Ultimately the outcomes of this work is discussed in Section 7.
 
\section{Painlev\'e Test:}
A partial differential equation is said to have Painlev\'e property if its solution is single valued about its movable singularity manifold \cite{weiss1983painleve}. So, if $\phi(x,y,t)=0$ is the singularity manifold of \eqref{1}, ($\phi(x,y,t)$ is analytic about the singularity manifold) and $u=u(x,y,t)$ is its solution then we assume that
\begin{eqnarray}
u=u(x,y,t)=\phi^{\alpha} \Sigma_{j=0}^{\infty}u_j\phi^j, \label{40}
\end{eqnarray}
where $u_j=u_j(x,y,t)$ and $\phi=\phi(x,y,t)$ are analytic near the singularity manifold $M=\left\lbrace(x,y,t):\phi(x,y,t)=0\right\rbrace.$

By leading order analysis we get $\alpha=-2$. Therefore,
\begin{eqnarray}
u=\Sigma_{j=0}^{\infty}u_j\phi^{j-2}.\label{41}
\end{eqnarray}
Now for the recursion relation we collect the coefficients of $\phi^{j-6}$ from \eqref{2} after substituting \eqref{41}. For $j=0$ we find that 
\begin{eqnarray}
u_0=-\frac{6b}{a}\phi_x \phi_t\label{42}
\end{eqnarray}
For $j\neq 0$ we obtain the recursion relation as
\begin{eqnarray}
& &(j+1)(j-4)(j-5)(j-6)bu_j\phi_{x}^3\phi_t=-[u_{j-4,xt}+(j-5)u_{j-3,x}\phi_t+(j-5)u_{j-3,t}\phi_x\nonumber\\& &+(j-4)(j-5)u_{j-2}\phi_x\phi_t+(j-5)u_{j-3}\phi_{xt}+u_{j-4,xx}+2(j-5)u_{j-3,x}\phi_x+(j-4)(j-5)u_{j-2}\phi_{x}^2\nonumber\\& &+(j-5)u_{j-3}\phi_{xx}+b\{ u_{j-4,xxxt}+(j-5)u_{j-3,xxx}\phi_t+3(j-5)u_{j-3,xxt}\phi_x+3(j-4)(j-5)u_{j-2,xx}\phi_{x}\phi_{t}\nonumber\\& & +3(j-5)u_{j-3,xx}\phi_{xt}+3(j-4)(j-5)u_{j-2,xt}\phi_{x}^2+3(j-3)(j-4)(j-5)u_{j-1,x}\phi_{x}^2\phi_t \nonumber\\& &+6(j-4)(j-5)u_{j-2,x}\phi_{x}\phi_{xt}+2(j-5)u_{j-3,xt}\phi_{xx}+3(j-4)(j-5)u_{j-2,x}\phi_{xx}\phi_t+3(j-5)u_{j-3,x}\phi_{xxt}\nonumber\\& &+(j-3)(j-4)(j-5)u_{j-1,t}\phi_{x}^3+3(j-3)(j-4)(j-5)u_{j-1}\phi_{x}^2\phi_{xt}+3(j-4)(j-5)u_{j-2,t}\phi_x \phi_{xx}\nonumber\\& &+3(j-3)(j-4)(j-5)u_{j-1}\phi_x \phi_{xx}\phi_{t}+3(j-4)(j-5)u_{j-2}\phi_{xt}\phi_{xx}+3(j-4)(j-5)u_{j-2}\phi_x \phi_{xxt}\nonumber\\& &+(j-5)u_{j-3,xt}\phi_{xx}+(j-5)u_{j-3,t}\phi_{xxx}+(j-4)(j-5)u_{j-2}\phi_{xxx}\phi_{t}+(j-5)u_{j-3}\phi_{xxxt} \}+k\{u_{j-4,yy}\nonumber\\& &+2(j-5)u_{j-3,y}\phi_y +(j-4)(j-5)u_{j-2}\phi_{y}^2+(j-5)u_{j-3}\phi_{yy} \}\nonumber\\& &+2a\Sigma_{r=1}^{j-1}\{u_{j-r,x}u_{r-2,x}+(j-r-2)u_{j-r}u_{r-1,x}\phi_x+(r-3)u_{j-r,x}u_{r-1}\phi_x+(j-r-2)(r-2)u_{j-r}u_r \phi_x^2\nonumber\\& &+u_{j-r}u_{r-2,xx}+2(r-3)u_{j-r}u_{r-1,x}\phi_{x}+(r-2)(r-3)u_{j-r}u_r\phi_x^2+(r-3)u_{j-r}u_{r-1}\phi_{xx} \}\nonumber\\& &+2a\{u_{0,x}u_{j-2,x}-2u_0u_{j-1,x}\phi_x+(j-3)u_{0,x}u_{j-1}\phi_x+u_0u_{j-2,xx}+2(j-3)u_0u_{j-1,x}\phi_x\nonumber\\& &+(j-3)u_0u_{j-1}\phi_{xx} \}]\label{43}
\end{eqnarray}
From the recursion relation \eqref{43} we find the resonances as $j=-1,4,5,6.$ the resonance at $j=-1$ indicates the arbitrariness of the singularity manifold. Resonances at $j=4,5,6$ indicates that $u_4,u_5,u_6$ must be arbitrary and the right hand side of \eqref{43} must have to vanish identically at $j=4,5,6.$ But by using the software Mathematica we found non zero right hand side for $j=4,5,6$. i.e the compatibility conditions for Painlev\'e test for these resonances are not satisfied. Thus the KP-BBM equation fails to pass the Painlev\'e test and hence this equation does not possess the Painlev\'e property.

\section{Lie Point Symmetry Analysis:}
Let us define the following point transformations:
$(x,y,t,u)\rightarrow(\hat{x},\hat{y}, \hat{t},\hat{u})$ where
\begin{eqnarray}
& &\hat{x}=\hat{x}(x,y,t,u(x,y,t);\epsilon),
\hat{y}=\hat{y}(x,y,t,u(x,y,t);\epsilon),
\hat{t}=\hat{t}(x,y,t,u(x,y,t);\epsilon),\nonumber\\& &
\hat{u}=\hat{u}(x,y,t,u(x,y,t);\epsilon).\label{17}
\end{eqnarray} 
Tangent vectors $(\xi,\gamma, \tau,\eta)$  at $(\hat{x},\hat{y},\hat{t},\hat{u})$ is defined by
\begin{eqnarray}
\frac{d\hat{x}}{d\epsilon}=\xi(\hat{x},\hat{y}, \hat{t},\hat{u}),\frac{d\hat{y}}{d\epsilon}=\gamma(\hat{x},\hat{y}, \hat{t},\hat{u}),\frac{d\hat{t}}{d\epsilon}=\tau(\hat{x},\hat{y},\hat{t},\hat{u}),\frac{d\hat{u}}{d\epsilon}=\eta(\hat{x},\hat{y},\hat{t},\hat{u})\label{18}
\end{eqnarray}
satisfying the conditions
\begin{eqnarray}
(\hat{x},\hat{y} ,\hat{t},\hat{u})|_{\epsilon=0}=(x,y,t,u)\label{19}
\end{eqnarray} 
Therefore, the Taylor's series of the Lie group action are 
\begin{eqnarray}
\hat{x}=x+\epsilon \xi(x,y,t,u)+o(\epsilon^2)\nonumber\\
\hat{y}=y+\epsilon \gamma(x,y,t,u)+o(\epsilon^2)\nonumber\\
\hat{t}=t+\epsilon \tau(x,y,t,u)+o(\epsilon^2)\nonumber\\
\hat{u}=u+\epsilon \eta(x,y,t,u)+o(\epsilon^2)\nonumber\\
\label{20}
\end{eqnarray}
The differential coefficients become:
\begin{eqnarray}
& &\hat{u}_{\hat{x}}=u_x+\epsilon\eta^x(x,y,t,u,u_t,u_y,u_x)+o(\epsilon^2)\nonumber\\& &
\hat{u}_{\hat{x}\hat{x}}=u_{xx}+\epsilon\eta^{xx}(x,y,t,u,u_t,u_y,u_x,...)+o(\epsilon^2)\nonumber\\& &
\hat{u}_{\hat{y}}=u_y+\epsilon\eta^y(x,y,t,u,u_t,u_y,u_x)+o(\epsilon^2)\nonumber\\& &
\hat{u}_{\hat{y}\hat{y}}=u_{yy}+\epsilon\eta^{yy}(x,y,t,u,u_t,u_y,u_x,...)+o(\epsilon^2)\nonumber\\& &
\hat{u}_{\hat{x}\hat{t}}=u_{xt}+\epsilon\eta^{xt}(x,y,t,u,u_t,u_y,u_x,...)+o(\epsilon^2)\nonumber\\& &
\hat{u}_{\hat{x}\hat{x}\hat{x}\hat{t}}=u_{xxxt}+\epsilon\eta^{xxxt}(x,y,t,u,u_t,u_y,u_x,...)+o(\epsilon^2)\nonumber\\
\label{21}
\end{eqnarray}
where
\begin{eqnarray}
& &\eta^x=D_x\eta-u_xD_x\xi-u_yD_x\gamma-u_tD_x\tau\nonumber\\& &
\eta^{xx}=D_x\eta^x-u_{xx}D_x\xi-u_{xy}D_x\gamma-u_{xt}D_x\tau\nonumber\\& &
\eta^y=D_y\eta-u_xD_y\xi-u_yD_y\gamma-u_tD_y\tau\nonumber\\& &
\eta^{yy}=D_y\eta^y-u_{xy}D_y\xi-u_{yy}D_y\gamma-u_{yt}D_y\tau\nonumber\\& &
\eta^{xt}=D_t\eta^x-u_{xx}D_t\xi-u_{xy}D_t\gamma-u_{xt}D_t\tau\nonumber\\& &
\eta^{xxxt}=D_t\eta^{xxx}-u_{xxxx}D_t\xi-u_{xxxy}D_t\gamma-u_{xxxt}D_t\tau\nonumber\\& &
\label{22}
\end{eqnarray}
and\\ $D_x=\frac{\partial}{\partial x}+u_x\frac{\partial}{\partial u}+u_{xx}\frac{\partial}{\partial u_x}+u_{xy}\frac{\partial}{\partial u_y}+u_{xt}\frac{\partial}{\partial u_t}+u_{xxt}\frac{\partial}{\partial u_{xt}}+...$ ,\\ $D_y=\frac{\partial}{\partial y}+u_y\frac{\partial}{\partial u}+u_{xy}\frac{\partial}{\partial u_x}+u_{yy}\frac{\partial}{\partial u_y}+u_{yt}\frac{\partial}{\partial u_t}+u_{xyt}\frac{\partial}{\partial u_{xt}}+...$ ,\\$D_t=\frac{\partial}{\partial t}+u_t\frac{\partial}{\partial u}+u_{xt}\frac{\partial}{\partial u_x}+u_{ty}\frac{\partial}{\partial u_y}+u_{tt}\frac{\partial}{\partial u_t}+u_{xtt}\frac{\partial}{\partial u_{xt}}+...$ .\\
The transformations \eqref{17} will be a symmetry transformation for \eqref{1} if it satisfies the following condition
\begin{eqnarray}
& &\hat{u}_{\hat{x}\hat{t}}+\hat{u}_{\hat{x}\hat{x}}+2a(\hat{u}_{\hat{x}})^2+2a\hat{u}\hat{u}_{\hat{x}\hat{x}}+b\hat{u}_{\hat{x}\hat{x}\hat{x}\hat{t}}+k\hat{u}_{\hat{y}\hat{y}}=0\nonumber\\
\Rightarrow & &u_{xt}+\epsilon\eta^{xt}+u_{xx}+\epsilon\eta^{xx}+2a(u_x+\epsilon\eta^{x})^2+2a(u+\epsilon\eta)(u_{xx}+\epsilon\eta^{xx})+b(u_{xxxt}+\epsilon\eta^{xxxt})\nonumber\\& &+k(u_{yy}+\epsilon\eta^{yy})=0\label{23}
\end{eqnarray}
Collecting the coefficients of $\epsilon$ from \eqref{23} we get 
\begin{eqnarray}
\eta^{xt}+\eta^{xx}+4au_x\eta^{x}+2au_{xx}\eta+2au\eta^{xx}+b\eta^{xxxt}+k\eta^{yy}=0,\label{24}
\end{eqnarray} 
which gives the symmetry condition (linearized) for \eqref{1}.

We calculate $\eta^{x},\eta^{xt},\eta^{xx}...$ etc from \eqref{22} and use them in \eqref{24}. The functions $\xi,\gamma,\tau,\eta$ are free from the terms $u_x, u_y, u_t, u_{xx}, u_{xy},......$  . So by equating the coefficients of the various derivative terms of $u$ to 0,  a set of equations is obtained, from which we derive the following
\begin{eqnarray}
\xi=c_1x+c_2, \gamma=c_1y+c_4,\tau=-2c_1t+c_3,\eta=c_1u+\frac{c_1}{2a};\label{25}
\end{eqnarray} 
 $c_1,\hspace{1mm} c_2,\hspace{1mm} c_3,\hspace{1mm} c_4$  are any constants.

The infinitesimal symmetry generator for \eqref{1} is  
\begin{eqnarray}
\Gamma & &=\xi\frac{\partial}{\partial x}+\gamma \frac{\partial}{\partial y}+\tau \frac{\partial}{\partial t}+\eta \frac{\partial}{\partial u}\nonumber\\& &=c_1\left(x\frac{\partial}{\partial x}+y\frac{\partial}{\partial y}-2t\frac{\partial}{\partial t}+(u+\frac{1}{2a})\frac{\partial}{\partial u}\right)+c_2\frac{\partial}{\partial x}+c_3\frac{\partial}{\partial t}+c_4\frac{\partial}{\partial y}\label{26}
\end{eqnarray}
Thus a Lie algebra $V$ is obtained which is generated by
\begin{eqnarray}
\Gamma_1=x\frac{\partial}{\partial x}+y\frac{\partial}{\partial y}-2t\frac{\partial}{\partial t}+\left(u+\frac{1}{2a}\right)\frac{\partial}{\partial u}, \Gamma_2=\frac{\partial}{\partial x},\Gamma_3=\frac{\partial}{\partial t},\Gamma_4=\frac{\partial}{\partial y}.\label{27}
\end{eqnarray}  
The commutator table is given below:

\begin{table}[H]
\begin{center}
\caption{Commutator table}
\begin{tabular}{|c|c|c|c|c|}
\hline
 \textbf{$[\Gamma_i ,\Gamma_j]$}& \textbf{$\Gamma_1$}&\textbf{$\Gamma_2$} &\textbf{$\Gamma_3$}&\textbf{$\Gamma_4$}\\
 \hline
 $\Gamma_1$& 0&-$\Gamma_2$&2$\Gamma_3$&$-\Gamma_4$\\
 $\Gamma_2$& $\Gamma_2$&0&0&0\\
 $\Gamma_3$&-2$\Gamma_3$&0&0&0\\
 $\Gamma_4$&$\Gamma_4$&0&0&0\\
 \hline

\end{tabular}
\end{center}
\end{table}
The first derived \cite{baumann2000symmetry} Lie algebra  $V^1$ is defined by $V^1=[V,V]=\{\Gamma_2,\Gamma_3,\Gamma_4\}$. Similarly the second derived Lie algebra is given by $V^2=[V^1,V^1]=0,$ which shows that the Lie algebra $V$ is solvable and hence the related differential equation can be solved. [A Lie algebra $V$ is solvable iff $V^{n+1}=[V^n, V^n]=0  $ for some natural $n$. ] 
\subsection{Construction of Adjoint matrix:}
To compute the adjoint relation we consider the Lie series \cite{hu2015direct}
\begin{eqnarray}
Ad_{exp(\epsilon \Gamma_i)} \Gamma_j=\Gamma_j-\epsilon [\Gamma_i ,\Gamma_j]+\frac{\epsilon^2}{2!} [\Gamma_i ,[\Gamma_i, \Gamma_j]]-\frac{\epsilon ^3}{3!}[\Gamma_i ,[\Gamma_i ,[\Gamma_i, \Gamma_j]]]+.... \label{161}
\end{eqnarray} 
The adjoint table is given below

\begin{table}[H]
\begin{center}
\caption{Adjoint table}
\begin{tabular}{|c|c|c|c|c|}
\hline
 \textbf{$Ad$}& \textbf{$\Gamma_1$}&\textbf{$\Gamma_2$} &\textbf{$\Gamma_3$}&\textbf{$\Gamma_4$}\\
 \hline
 $\Gamma_1$& $\Gamma_1$&$e^\epsilon \Gamma_2$&$e^{-2\epsilon} \Gamma_3$& $e^\epsilon \Gamma_4$\\
 $\Gamma_2$& $(\Gamma_1-\epsilon \Gamma_2)$& $\Gamma_2$&$\Gamma_3$&$\Gamma_4$\\
 $\Gamma_3$&$(\Gamma_1+2\epsilon \Gamma_3)$&$ \Gamma_2$&$\Gamma_3$&$\Gamma_4$\\
 $\Gamma_4$&$(\Gamma_1-\epsilon \Gamma_4)$&$ \Gamma_2$&$\Gamma_3$&$\Gamma_4$\\
 \hline

\end{tabular}
\end{center}
\end{table}
The adjoint action of $\Gamma_1$ on any vector $\Gamma \in V$ is given by
\begin{eqnarray}
Ad_{exp(\epsilon \Gamma_1)} \Gamma =d_1 \Gamma_1+d_2 \Gamma_2+d_3 \Gamma_3+d_4 \Gamma_4, \label{162}
\end{eqnarray}
which can be expressed in the matrix form as
\begin{eqnarray}
Ad_{exp(\epsilon \Gamma_1)} \Gamma =[d_1, d_2, d_3, d_4] A_1 [\Gamma_1, \Gamma_2, \Gamma_3, \Gamma_4].\nonumber
\end{eqnarray}
where
\begin{eqnarray}
A_1 =\begin{bmatrix}
1&0&0&0\\0&e^{\epsilon_1}&0&0\\0&0&e^{-2\epsilon_1}&0\\0&0&0&e^{\epsilon_1}
\end{bmatrix}
\nonumber
\end{eqnarray}
Similarly we obtain
\begin{eqnarray}
A_2=\begin{bmatrix}
1&-\epsilon_2 &0&0\\0&1&0&0\\0&0&1&0\\0&0&0&1\nonumber \\
\end{bmatrix}
,A_3=\begin{bmatrix}
1&0&2\epsilon_3 &0\\0&1&0&0\\0&0&1&0\\0&0&0&1\nonumber \\
\end{bmatrix}
,A_4=\begin{bmatrix}
1&0&0 &-\epsilon_4 \\0&1&0&0\\0&0&1&0\\0&0&0&1\nonumber \\
\end{bmatrix}
\end{eqnarray}
the global adjoint matrix $A$ is given by
\begin{eqnarray} 
A=\begin{bmatrix}
1&-\epsilon_2 &2\epsilon_3 &-\epsilon_4\\0&e^{\epsilon_1}&0&0\\0&0&e^{-2\epsilon_1}&0\\0&0&0&e^{\epsilon_1}
\end{bmatrix}.\label{163}
\end{eqnarray}
The general adjoint transformation equation is given by
\begin{eqnarray}
(\tilde{a}_1,\tilde{a}_2,\tilde{a}_3,\tilde{a}_4)=(a_1,a_2,a_3,a_4)A, \label{170}
\end{eqnarray}
by which the vector $a_1\Gamma_1+a_2\Gamma_2+a_3\Gamma_3+a_4\Gamma_4$ is transformed into the vector $\tilde{a}_1\Gamma_1+\tilde{a}_2\Gamma_2+\tilde{a}_3\Gamma_3+\tilde{a}_4\Gamma_4$ by adjoint action.
\subsection{Calculation of the Invariants of the Lie Algebra}
In this section we are going to find the invariants of the obtained Lie algebra by a method proposed by Hu et. al. in \cite{hu2015direct}.
Let $U, W$ be two elements of the Lie algebra spanned by ${\Gamma_1,\Gamma_2,\Gamma_3,\Gamma_4}$. Then $U,W$ can be expressed as $U=\Sigma_{i=1}^4 a_i \Gamma_i$ and $W=\Sigma_{j=1}^4 b_j \Gamma_j$.

Now
\begin{eqnarray}
Ad_{exp(\epsilon W)}U& &=U-\epsilon [W ,U]+\frac{\epsilon^2}{2!} [W ,[W, U]]-....\nonumber \\ & &=U-\epsilon [W ,U]+o(\epsilon^2)\nonumber \\ & &=(a_1 \Gamma_1+a_2 \Gamma_2+a_3 \Gamma_3+a_4 \Gamma_4)\nonumber \\& & \hspace{1cm}-\epsilon[(b_1 \Gamma_1+b_2 \Gamma_2+b_3 \Gamma_3+b_4 \Gamma_4),(a_1 \Gamma_1+a_2 \Gamma_2+a_3 \Gamma_3+a_4 \Gamma_4)]+o(\epsilon^2)\nonumber \\ & &=(a_1 \Gamma_1+a_2 \Gamma_2+a_3 \Gamma_3+a_4 \Gamma_4)-\epsilon ((\theta_1 \Gamma_1+\theta_2 \Gamma_2+\theta_3 \Gamma_3+\theta_4 \Gamma_4))+o(\epsilon^2),\nonumber \\ \label{164}
\end{eqnarray}
where $\theta_i(a_1,...,a_4,b_1,...,b_4)$ can easily be obtained from the commutator table. Thus the action of $Ad_{exp(\epsilon W)}$ on $U$ can be written as

 $(a_1,a_2,a_3,a_4)\rightarrow (a_1-\epsilon \theta_1,a_2-\epsilon \theta_2,a_3-\epsilon \theta_3,a_4-\epsilon \theta_4)+o(\epsilon^2)$.\\ 
Let $\phi$ be an invariant of the Lie algebra. Then 
\begin{eqnarray}
\phi (a_1,a_2,a_3,a_4)=\phi (a_1-\epsilon \theta_1 +o(\epsilon^2),a_2-\epsilon \theta_2+o(\epsilon^2),a_3-\epsilon \theta_3+o(\epsilon^2),a_4-\epsilon \theta_4+o(\epsilon^2))\label{165}
\end{eqnarray} 
By applying Taylor's theorem, the right hand side of \eqref{165} becomes
\begin{eqnarray}
& &\phi (a_1-\epsilon \theta_1 +o(\epsilon^2),a_2-\epsilon \theta_2+o(\epsilon^2),a_3-\epsilon \theta_3+o(\epsilon^2),a_4-\epsilon \theta_4+o(\epsilon^2))=\phi (a_1,a_2,a_3,a_4)\nonumber\\ & &
\hspace{5cm} -\epsilon \left(\theta_1 \frac{\partial \phi}{\partial a_1}+\theta_2 \frac{\partial \phi}{\partial a_2}+\theta_3 \frac{\partial \phi}{\partial a_3}+\theta_4 \frac{\partial \phi}{\partial a_4} \right)+o(\epsilon^2).\label{166}
\end{eqnarray}
Since $\phi$ is an invariant we must have
\begin{eqnarray}
\theta_1 \frac{\partial \phi}{\partial a_1}+\theta_2 \frac{\partial \phi}{\partial a_2}+\theta_3 \frac{\partial \phi}{\partial a_3}+\theta_4 \frac{\partial \phi}{\partial a_4}=0. \label{167}
\end{eqnarray}
We calculate $\theta_i$ from the commutator table as follows

$\theta_1=0, \hspace{2mm}\theta_2=-b_1a_2+b_2a_1,\hspace{2mm}\theta_3=2b_1a_3-2b_3a_1,\hspace{2mm}\theta_4=-b_1a_4+b_4a_1$.

Then \eqref{167} becomes
\begin{eqnarray}
(b_2a_1-b_1a_2)\frac{\partial \phi}{\partial a_2}+2(b_1a_3-b_3a_1)\frac{\partial \phi}{\partial a_3}+(b_4a_1-b_1a_4)\frac{\partial \phi}{\partial a_4}=0\label{168}
\end{eqnarray}
Equating the coefficients of each $b_i$ to 0, we get the following system of partial differential equations
\begin{eqnarray}
& &a_1\frac{\partial \phi}{\partial a_4}=0,\nonumber\\& &
a_1\frac{\partial \phi}{\partial a_3}=0,\nonumber\\& &
a_1\frac{\partial \phi}{\partial a_2}=0,\nonumber\\& &
a_2\frac{\partial \phi}{\partial a_2}-2a_3\frac{\partial \phi}{\partial a_3}+a_4\frac{\partial \phi}{\partial a_4}=0,\nonumber\\\label{169}
\end{eqnarray}
Solving the above system we get $\phi(a_1,a_2,a_3,a_4)=F(a_1)$, an arbitrary function of $a_1$. 
\subsection{One Dimensional Optimal System of Lie Sub-algebras}
In this section, we are looking for a one dimensional optimal system of subalgebras of the Lie algebra $V$ by using Hu-Li-Chen algorithm \cite{hu2015direct}. A list of one dimensional subalgebras is said to form an one dimensional optimal system if each subalgebra of the Lie algebra is equivalent to a definite member of this list \cite{saha2020invariant}.

 According to the Hu-Li-Chen algorithm at first the basic invariant(s) is (are) to be scaled to '-1,1,0' or '1,0' respectively as they are even or odd. Then for each case one representative element $\tilde{v}=\tilde{a}_1\Gamma_1+\tilde{a}_2\Gamma_2+\tilde{a}_3\Gamma_3+\tilde{a}_4\Gamma_4$ has to be selected for which the adjoint transformation equation \eqref{170} can be solved for $\epsilon_1,\epsilon_2,\epsilon_3,\epsilon_4$. At the time of scaling it must be kept in mind that once an invariant is scaled to a nonzero integer then the other invariants (if there be any) can not be adjusted.
 
 By using \eqref{163} in \eqref{170}, the adjoint transformation equation can be explicitly written as
\begin{eqnarray}
 (\tilde{a}_1,\tilde{a}_2,\tilde{a}_3,\tilde{a}_4)=(a_1,-\epsilon_2 a_1+e^{\epsilon_1}a_2,2\epsilon_3 a_1+e^{-2\epsilon_1}a_3,-\epsilon_4 a_1+e^{\epsilon_1}a_4). \label{171}
\end{eqnarray}  
The invariant of this Lie algebra is found to be $\phi(a_1,a_2,a_3,a_4)=F(a_1)$, where $F$ is an arbitrary function of $a_1$. $a_1$ is a basic invariant whose degree is one, an odd integer. So, by Hu-Li-Chen algorithm we have to consider two cases: $a_1=1$ and $a_1=0$.\\
\textbf{\underline{Case 1:} $a_1=1$}\\
Let us choose the representative element as $\tilde{v}=\Gamma_1$ i.e. $\tilde{a}_2=\tilde{a}_3=\tilde{a}_4=0$ and $\tilde{a}_1=1$.\\Under such choice the adjoint transformation equation \eqref{171} can be solved for $\epsilon_i$ $(i=1,2,3,4)$ as $\epsilon_1=0$, $\epsilon_2=a_2$, $\epsilon_3=-\frac{1}{2}a_3$, $\epsilon_4=a_4$ which indicates that the selected representative element is correct. Therefore, all the elements of the form $\Gamma_1+a_2\Gamma_2+a_3\Gamma_3+a_4\Gamma_4$ are equivalent to $\Gamma_1$.\\
\textbf{\underline{Case 2:} $a_1=0$}\\
For $a_1=0$, the system of equations \eqref{169} reduces to $$a_2\frac{\partial \phi}{\partial a_2}-2a_3\frac{\partial \phi}{\partial a_3}+a_4\frac{\partial \phi}{\partial a_4}=0$$ solving which we get $$\phi(a_1,a_2,a_3,a_4)=F(a_2^2a_3,a_4^2a_3)$$ where $F$ is an arbitrary function of two variables. Thus we have two basic invariants $\Delta_1=a_2^2a_3$ and $\Delta_2=a_4^2a_3$ both of which are odd degree polynomials. According to Hu-Li-Chen algorithm we have three sub-cases $$\{\Delta_1=1, \Delta_2=c\},\hspace{3mm}\{\Delta_1=0, \Delta_2=1\},\hspace{3mm}\{\Delta_1=0, \Delta_2=0\}$$\\

\textbf{\underline{Subcase 2.1}:} $\{\Delta_1=a_2^2a_3=1, \Delta_2=a_4^2a_3=c\}$

Obviously $a_2\neq 0$, $a_3>0$. The following subcases under this subcase 2.1 are to be considered:\\

\textbf{Subsubcase 2.1.1:} $a_2>0$, $a_3>0$. Then from \eqref{171} we have $$\tilde{a}_1=0,\hspace{2mm} e^{\epsilon_1}a_2=\tilde{a}_2,\hspace{2mm}e^{-2\epsilon_1}a_3=\tilde{a}_3,\hspace{2mm}e^{\epsilon_1}a_4=\tilde{a}_4$$ which give $$\epsilon_1=\log \frac{\tilde{a}_2}{a_2}=-\frac{1}{2}\log \frac{\tilde{a}_3}{a_3}=\log \frac{\tilde{a}_4}{a_4},\hspace{3mm} \tilde{a}_2>0,\tilde{a}_3>0$$ which implies $$\tilde{a}_2^2\tilde{a}_3=a_2^2a_3=1,\hspace{3mm}\tilde{a}_4^2\tilde{a}_3=a_4^2a_3=c$$
Considering all of the above restrictions on $\tilde{a}_1,\tilde{a}_2,\tilde{a}_3,\tilde{a}_4$ we can take$$\tilde{a}_1=0,\tilde{a}_2=1,\tilde{a}_3=1,\tilde{a}_4=\sqrt{c}$$
Therefore, the corresponding representative element is $$\Gamma_2+\Gamma_3+\sqrt{c}\Gamma_4$$.

\textbf{Subsubcase 2.1.2:} $a_2<0$, $a_3>0$. In this case the representative element can be chosen as $$-\Gamma_2+\Gamma_3+\sqrt{c}\Gamma_4$$\\

\textbf{\underline{Subcase 2.2}:} $\{\Delta_1=a_2^2a_3=0, \Delta_2=a_4^2a_3=1\}$

Clearly here $a_2=0,\hspace{1mm} a_3>0$ and either $a_4>0$ or $a_4<0$. Thus we have to consider two subcases under this subcase 2.2:

\textbf{Subsubcase 2.2.1:} $a_2=0$, $a_3>0$, $a_4>0$. Proceeding in similar fashion we found the representative $\Gamma_3+\Gamma_4$.

\textbf{Subsubcase 2.2.2:} $a_2=0$, $a_3>0$, $a_4<0$. The corresponding representative is found to be $\Gamma_3-\Gamma_4$.\\

\textbf{\underline{Subcase 2.3}:} $\{\Delta_1=a_2^2a_3=0, \Delta_2=a_4^2a_3=0\}$ which can be divided into following subcases:\\

\textbf{Subsubcase 2.3.1:} $a_2=0$, $a_3\neq 0$, $a_4=0$. There are two possibilities $a_3>0$ and $a_3<0$, correspondingly representative elements are found to be $\Gamma_3$ and $-\Gamma_3$ which are equivalent to each other. So in this Subsubcase 2.3.1 we choose $\Gamma_3$ as representative member in the optimal list.\\  

\textbf{Subsubcase 2.3.2:} $a_3=0$. All the possibilities and the corresponding representatives are listed below:\\
$(i)$  $a_1=0$, $a_2> 0$, $a_3=0$, $a_4=0$ $\Rightarrow$ representative $\Gamma_2$\\
$(ii)$  $a_1=0$, $a_2< 0$, $a_3=0$, $a_4=0$ $\Rightarrow$ representative $-\Gamma_2$\\
$(iii)$  $a_1=0$, $a_2= 0$, $a_3=0$, $a_4>0$ $\Rightarrow$ representative $\Gamma_4$\\
$(iv)$  $a_1=0$, $a_2= 0$, $a_3=0$, $a_4<0$ $\Rightarrow$ representative $-\Gamma_4$\\
$(v)$  $a_1=0$, $a_2> 0$, $a_3=0$, $a_4>0$ $\Rightarrow$ representative $\Gamma_2+\Gamma_4$\\
$(vi)$  $a_1=0$, $a_2> 0$, $a_3=0$, $a_4<0$ $\Rightarrow$ representative $\Gamma_2-\Gamma_4$\\
$(vii)$  $a_1=0$, $a_2< 0$, $a_3=0$, $a_4>0$ $\Rightarrow$ representative $-\Gamma_2+\Gamma_4$\\
$(viii)$  $a_1=0$, $a_2< 0$, $a_3=0$, $a_4<0$ $\Rightarrow$ representative $\Gamma_2-\Gamma_4$\\

Thus in this case we get four distinct inequivalent representative vectors as$$\Gamma_2,\hspace{2mm}\Gamma_4,\hspace{2mm}\Gamma_2+\Gamma_4,\hspace{2mm}\Gamma_2-\Gamma_4$$.

Therefore, by listing all the above selected representatives we have the following optimal system of sub-algebras:$$\{\Gamma_1\},\hspace{2mm}\{\Gamma_2\},\hspace{2mm}\{\Gamma_3\},\hspace{2mm}\{\Gamma_4\},\hspace{2mm}\{\Gamma_2+\Gamma_4\},\hspace{2mm}\{\Gamma_2-\Gamma_4\},\hspace{2mm}\{\Gamma_3+\Gamma_4\},\hspace{2mm}\{\Gamma_3-\Gamma_4\},$$ $$\hspace{2mm}\{\Gamma_2+\Gamma_3+\sqrt{c} \Gamma_4\},\hspace{2mm}\{-\Gamma_2+\Gamma_3+\sqrt{c} \Gamma_4\}$$
\subsection{Similarity Reductions and Exact solutions}
The invariant surface condition for \eqref{1} is given by
\begin{eqnarray}
\eta -\xi \frac{\partial u}{\partial x}-\gamma \frac{\partial u}{\partial y}-\tau \frac{\partial u}{\partial t}=0, \label{150}
\end{eqnarray}
which is a quasilinear equation in $u(x,y,t)$ and its auxiliary equations are
\begin{eqnarray}
\frac{dx}{\xi}=\frac{dy}{\gamma}=\frac{dt}{\tau}=\frac{du}{\eta}\label{151}
\end{eqnarray}
Let us find the similarity reduction corresponding to the subalgebra $\{\Gamma_2+\Gamma_3+\sqrt{c} \Gamma_4\}$ with taking the constant $c=1$. The auxiliary equation \eqref{151} becomes
\begin{eqnarray}
\frac{dx}{1}=\frac{dy}{1}=\frac{dt}{1}=\frac{du}{0}.\nonumber
\end{eqnarray}
 We find the general solution of \eqref{150} in the following forms:
\begin{eqnarray}
u(x,y,t)=F(x-y,y-t)\label{152}\\u(x,y,t)=G(x-y,x-t)\label{153} \\u(x,y,t)=H(x-t,y-t)\label{154}
\end{eqnarray} 
By putting \eqref{152} in \eqref{1} we get the following similarity reduction with the similarity variables $\alpha=x-y, \beta=y-t$ 
\begin{eqnarray}
-F_{\alpha\beta}+F_{\alpha\alpha}+a(F^2)_{\alpha\alpha}-bF_{\alpha\alpha\alpha\beta}+k(F_{\alpha\alpha}-2F_{\alpha\beta}+F_{\beta\beta})=0,\label{155}
\end{eqnarray}
solving which we get the following exact solution of \eqref{1}
\begin{eqnarray}
u(x,y,t)=\frac{3(\lambda-1)(k\lambda-k-1)}{2a}Sech^2 \left\lbrace \frac{1}{2}\sqrt{\frac{(\lambda-1)(k\lambda-k-1)}{b\lambda}}(x+(\lambda-1)y-\lambda t)\right\rbrace,\label{156}
\end{eqnarray}
where $\lambda$ is an arbitrary constant.

By putting \eqref{153} in \eqref{1} we get another similarity reduction with the similarity variables $\alpha=x-y, \beta=x-t$ 
\begin{eqnarray}
(1+k)G_{\alpha\alpha}+G_{\alpha \beta}+a(G^2)_{\alpha\alpha}+2a(G^2)_{\alpha\beta}+a(G^2)_{\beta\beta}-bG_{\alpha\alpha\alpha\beta}-3bG_{\alpha\alpha\beta\beta}
-3bG_{\alpha\beta\beta\beta}-bG_{\beta\beta\beta\beta}=0,\nonumber\\ \label{157}
\end{eqnarray}
solving which we get the following exact solution of \eqref{1}
\begin{eqnarray}
u(x,y,t)=-\frac{3(2+k+\lambda-\lambda^2)}{2a(1+\lambda)^2}Sech^2 \left\lbrace \frac{1}{2}\sqrt{\frac{2+k+\lambda-\lambda^2}{b\lambda(1+\lambda)^3}}((1+\lambda)x-y-\lambda t)\right
\rbrace, \label{158}
\end{eqnarray}
where $\lambda (\neq -1)$ is any constant other than -1.

Now by putting \eqref{154} in \eqref{1} we get one another similarity reduction with the similarity variables $\alpha=x-t, \beta=y-t$

\begin{eqnarray}
kH_{\beta\beta}-H_{\alpha\beta}+a(H^2)_{\alpha\alpha}-bH_{\alpha\alpha\alpha\alpha}-bH_{\alpha\alpha\alpha\beta}=0, \label{159}
\end{eqnarray}
solving which we get another exact solution of \eqref{1} as
\begin{eqnarray}
u(x,y,t)=\frac{3(k\lambda^2-\lambda)}{2a}Sech^2 \left\lbrace \frac{1}{2} \sqrt{\frac{k\lambda^2-\lambda}{b(\lambda+1)}}(x+\lambda y-(\lambda+1)t)\right\rbrace, \label{160}
\end{eqnarray}
where $\lambda (\neq -1)$ is an any constant other than -1.

Amplitude of each solitary wave \eqref{156}, \eqref{158}, \eqref{160} decreases with respect to the nonlinear coefficient $a$ and the widths increase with respect to the dispersion coefficient $b$. $a$, $b$ don't have any effect on the velocity of the above solitary waves.
\begin{figure}[H]
\includegraphics[width=8cm]{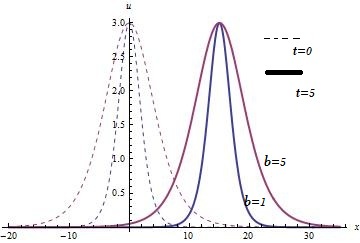}
\caption{Effect of the dispersion coefficient $b$ on the width and velocity of the solitary wave \eqref{160}, taking $k=1, \lambda=2, a=1, y=0$.}
\end{figure}
The similarity reductions and corresponding exact solutions for each member of the optimal system can be obtained similarly.
\section{Solution by Homogeneous Balance Method:}
Here we are going to solve \eqref{1} by considering 
\begin{eqnarray}
u(x,y,t)=\frac{\partial ^p f(\phi(x,y,t))}{\partial x^p}+u_1(x,y,t),\label{3}
\end{eqnarray}
in which $\phi,u_1$, $f$ are functions of $x,y,t$ to be found. We choose $\phi$ in such a way that its $x-$derivatives and $t-$derivatives are same. i.e. $\phi_x=\phi_t, \phi_{xt}=\phi_{xx}, \phi_{xxxt}=\phi_{xxxx}$ etc. 

Balancing the highest power of $\phi_x$ we find that $p=2$. Therefore,
\begin{eqnarray}
u(x,y,t)=\frac{\partial ^2 f(\phi(x,y,t))}{\partial x^2}+u_1(x,y,t)=\frac{d^2f}{d\phi^2}\left( \frac{d\phi}{dx}\right)^2+\frac{df}{d\phi}\frac{d^2\phi}{dx^2}+u_1,\label{4}
\end{eqnarray} 

Using \eqref{4} in \eqref{1} we get,
\begin{eqnarray}
& &(2af'''^2+2af''f^{iv}+bf^{vi})\phi_x^6+(24af''f'''+2af'f^{iv}+15bf^v)\phi_x^4\phi_{xx}+(2f^{iv}\phi_x^4+24af''^2\phi_x^2\phi_{xx}^2\nonumber\\& &+4af'f'''\phi_x^3\phi_{xxx}+8af''^2\phi_x^3\phi_{xxx}+12af'f'''\phi_x^2\phi_{xx}^2+2af^{iv}\phi_x^4u_1+45bf^{iv}\phi_x^2\phi_{xx}^2+20bf^{iv}\phi_x^3\phi_{xxx}\nonumber\\& &+kf^{iv}\phi_x^2\phi_y^2)+(12f'''\phi_x^2\phi_{xx}+20af'f''\phi_x\phi_{xx}\phi_{xxx}+4af'''\phi_x^3u_{1x}+2af'f''\phi_x^2\phi_{xxxx}+6af'f''\phi_{xx}^3\nonumber\\& &+12af'''\phi_x^2\phi_{xx}u_1+15bf'''\phi_{xx}^3+60bf'''\phi_x\phi_{xx}\phi_{xxx}+15bf'''\phi_x^2\phi_{xxxx}+4kf'''\phi_x\phi_{xy}\phi_y+kf'''\phi_x^2\phi_{yy}\nonumber\\& &+kf'''\phi_{xx}\phi_y^2)+(6f''\phi_{xx}^2+8f''\phi_x\phi_{xxx}+2af'^2\phi_{xxx}^2+12af''\phi_x\phi_{xx}u_{1x}+2af''\phi_x^2u_{1xx}\nonumber\\& &+2af'^2\phi_{xx}\phi_{xxxx}+6af''\phi_{xx}^2u_1+8af''\phi_x\phi_{xxx}u_1+10bf''\phi_{xxx}^2+15bf''\phi_{xx}\phi_{xxxx}+6bf''\phi_x\phi_{xxxxx}\nonumber\\& &+2kf''\phi_{xy}^2+2kf''\phi_x\phi_{xyy}+2kf''\phi_{xxy}\phi_y+kf''\phi_{xx}\phi_{yy})+(2f'\phi_{xxxx}+4af'\phi_{xxx}u_{1x}+2af'\phi_{xx}u_{1xx}\nonumber\\& &+2af'\phi_{xxxx}u_1+bf'\phi_{xxxxxx}+kf'\phi_{xxyy})+(u_{1xt}+u_{1xx}+2au_{1x}^2+2au_1u_{1xx}+bu_{1xxxt}+ku_{1yy})\nonumber\\& &=0,\label{5}
\end{eqnarray}
where $f'=\frac{df}{d\phi}, f''=\frac{d^2f}{d\phi^2}, ... ...$  .

Equating coefficient of $\phi_x^6$ in \eqref{5} to zero we get
\begin{eqnarray}
2af'''^2+2af''f^{iv}+bf^{vi}=0\label{6}
\end{eqnarray} 
$f=ln\phi$ is a solution of \eqref{6} if $a=6b.$
So, let us take 
\begin{eqnarray}
& &f=ln\phi\label{7}\\& &a=6b\label{8}
\end{eqnarray}
Then $f''f'''=-\frac{1}{12}f^v$, $f'f^{iv}=-\frac{1}{4}f^v$, $f''^2=-\frac{1}{6}f^{iv}$, $f'f'''=-\frac{1}{3}f^{iv}$, $f'f''=-\frac{1}{2}f'''$, $f'^2=-f''.$
Using the above results in \eqref{5} and then equating the coefficients of $f^{iv}, f''', f'', f' $ and $f$ free terms to zero we get a system of differential equations in $\phi$, which has a solution of the form 
\begin{eqnarray}
\phi(x,y,t)=1+exp(\alpha x+\beta y+ \alpha t+\theta_0).\label{9}
\end{eqnarray}
[remember that  $x-$derivatives and $t-$derivatives of $\phi$ are same].\\
Using \eqref{7}, \eqref{8}, \eqref{9} in \eqref{5} we obtain the following system of algebraic equations:
\begin{eqnarray}
& &2\alpha^4+12b\alpha^4u_1+b\alpha^6+k\alpha^2\beta^2=0,\label{10}\\& &
12\alpha^4+72b\alpha^4u_1+6b\alpha^6+6k\alpha^2\beta^2+24b\alpha^3u_{1x}=0,\label{11}\\& &
14\alpha^4+84b\alpha^4u_1+7b\alpha^6+7k\alpha^2\beta^2+72b\alpha^3u_{1x}+12b\alpha^2u_{1xx}=0,\label{12}\\& &
2\alpha^4+12b\alpha^4u_1+b\alpha^6+k\alpha^2\beta^2+24b\alpha^3u_{1x}+12b\alpha^2u_{1xx}=0,\label{13}\\& &
u_{1xt}+u_{1xx}+2au_{1x}^2+2au_1u_{1xx}+bu_{1xxxt}+ku_{1yy}=0.\label{14}
\end{eqnarray}
From \eqref{10} we see that $u_1$ is a constant and 
\begin{eqnarray}
\beta=\pm \sqrt{-\frac{2\alpha^2+12b\alpha^2u_1+b\alpha^4}{k}},\label{15}
\end{eqnarray}
Using \eqref{7}, \eqref{9} in \eqref{4} we find that
\begin{eqnarray}
u(x,y,t)=\frac{\alpha^2}{2+2cosh(\alpha x+\beta y+\alpha t+\theta_0)}+u_1\label{16}
\end{eqnarray}
gives a solution of \eqref{1} with $a=6b$; where $\beta$ is given by \eqref{15} and $\alpha, u_1, \theta_0$ are arbitrary constants.
\begin{figure}[H]
\includegraphics[width=8cm]{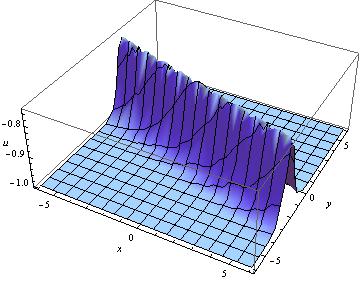}{t=0}
\includegraphics[width=8cm]{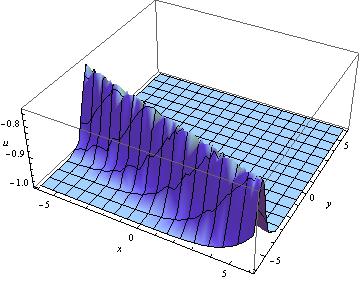}{t=10}
\includegraphics[width=8cm]{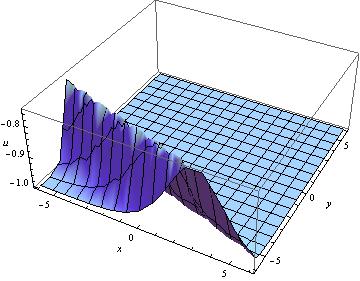}{t=15}
\caption{Movement of the wave \eqref{16} with $\alpha=\gamma =k=1, b=1, u_1 =-1,\theta_{0}$}
\end{figure}
\section{Solution by Tanh method:}
To solve equation \eqref{1} by Tanh method, we transform the partial differential equation \eqref{1} to the following ordinary differential equation by using the transformation $u(x,y,t)=f(z)$ where $z=x-\lambda y-\omega t$:
\begin{eqnarray}
& &(1-\omega+k\lambda^2)f''+a(f^2)''-b\omega f^{iv}=0\nonumber\\ \Rightarrow & &(1-\omega+k\lambda^2)f+af^2-b\omega f''=0.\label{28}
\end{eqnarray}
(integrating twice and taking the integration constant 0).\\
Let us impose the following boundary condition on \eqref{28}
\begin{eqnarray}
f(z)\rightarrow 0 \hspace{5mm}and \hspace{5mm}f^n(z)\rightarrow 0 \hspace{5mm}(n\in \mathbb{N} ) \hspace{5mm}as \hspace{5mm}z\rightarrow \pm \infty, \label{45}
\end{eqnarray}
$f^n(z)$ denotes the $n$-th derivative of $f$ with respect to $z$.

For finding the solution in Tanh method we suppose that
\begin{eqnarray}
u(x,y,t)=f(z)=S(Y)=\Sigma_{n=0}^J c_nY^n, \label{46}
\end{eqnarray}
where $Y=tanh \hspace{1mm} z=tanh\hspace{1mm} (x-\lambda y-\omega t)$.
Then
\begin{eqnarray}
& &f'(z)=\frac{df}{dz}=\frac{df}{dY} \frac{dY}{dz}=(1-Y^2)\frac{dS}{dY},\label{47}\\& &
f''(z)=\frac{d^2f}{dz^2}=\frac{d}{dz} \left(\frac{df}{dz}\right)=\frac{d}{dY}\left((1-Y^2)\frac{dS}{dY}\right)\frac{dY}{dz}\nonumber\\& &\hspace{11mm}=-2Y(1-Y^2)\frac{dS}{dY}+(1-Y^2)^2\frac{d^2S}{dY^2}\label{48}
\end{eqnarray}
By using \eqref{47}, \eqref{48} equation \eqref{28} becomes:
\begin{eqnarray}
(1-\omega+k\lambda^2)S(Y)+aS^2(Y)+2b\omega Y(1-Y^2)\frac{dS}{dY}-b\omega (1-Y^2)^2\frac{d^2S}{dY^2}=0\label{49}
\end{eqnarray}
By leading order analysis from \eqref{46} and \eqref{49} it is found that $J=2$.\\
The boundary conditions \eqref{45} becomes
\begin{eqnarray}
S(Y)\rightarrow 0 \hspace{3mm} as \hspace{3mm} Y\rightarrow \pm 1.
\end{eqnarray}
Considering the case $Y\rightarrow 1$, the solution takes the following form
\begin{eqnarray}
S(Y)=d_0(1-Y)(1+d_1Y)\label{50}
\end{eqnarray}  
We directly substitute \eqref{50} in \eqref{49} and make equal the coefficients of various degrees of $Y$ to 0. A set of algebraic equations is thus obtained, solving which we get 
\begin{eqnarray}
d_0=-\frac{6b \omega}{a}, \hspace{2mm} d_1=1, \hspace{2mm} \omega=\frac{1+k \lambda^2}{1+4b},\hspace{2mm}(b\neq-\frac{1}{4}).
\end{eqnarray} 
Thus from \eqref{46} and \eqref{50} we obtain a solitary wave solution of \eqref{1} with $b\neq -\frac{1}{4}$ as
\begin{eqnarray}
u(x,y,t)=-\frac{6b(1+k\lambda ^2)}{a(1+4b)}Sech^2 \left(x-\lambda y-\frac{1+k\lambda ^2}{1+4b}t \right),\label{51}
\end{eqnarray}
where $\lambda$ is an arbitrary constant.

The amplitude decreases with respect to the nonlinear coefficient $a$, though  $a$ has no effect on the velocity. The amplitude increases and the velocity decreases with respect to the dispersion coefficient $b$. That is the waves with greater amplitude moves slower than the waves with less amplitude. Though amplitude increases with respect to $b$, it has an upper limit $\frac{3(1+k\lambda^2)}{2a}$ as $b \rightarrow \infty$. 
\begin{figure}[H]
\includegraphics[width=8cm]{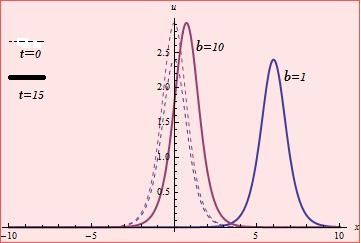}
\caption{Effect of the dispersion coefficient $b$ on the amplitude and velocity of the solitary wave \eqref{51}, taking $k=1, \lambda=1, a=-1, y=0$.}
\end{figure} 
\begin{figure}[H]
\includegraphics[width=8cm]{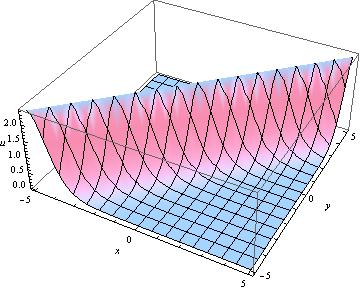}{t=0}
\includegraphics[width=8cm]{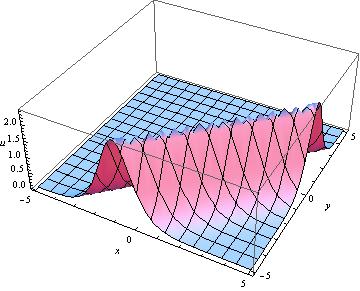}{t=10}
\includegraphics[width=8cm]{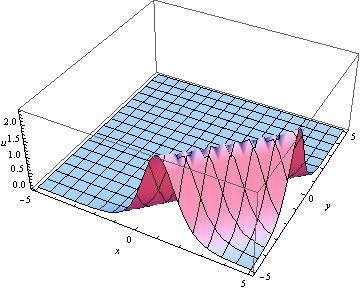}{t=15}
\caption{Movement of the solitary wave \eqref{51} with $\lambda=b=k=1, a=-1$}
\end{figure}
\section{Conclusion:}
This research article is aimed to study the Painlev\'e property, Lie point symmetry analysis and finding exact solutions of the KP-BBM equation. Painlev\'e test is performed by using WTC method \cite{weiss1983painleve} and the equation is found to fail the Painlev\'e test which indicates its nonintegrability. The Lie point symmetry generators are obtained explicitly and the obtained Lie algebra is found to be solvable. The general adjoint transformation matrix, invariants of the Lie algebra are obtained and one dimensional optimal system for subalgebras is constructed by Hu-Li-Chen algorithm. Three similarity reductions are obtained and solved analytically. We got three solitary wave solutions from these similarity reductions and it is seen that the amplitudes of these solitary waves are inversely proportional with the nonlinear parameter $a$ and the dispersion parameter $b$ increases their widths. We successfully apply HB method and $Tanh$ method  to derive exact solutions, that once again  prove the efficiency of these methods. Solitary wave like solution are obtained by the methods of HB and $Tanh$ and are sketched with suitable  parametric values. Figure 2 and Figure 4 depict the time evolution of the solitary waves in \eqref{16} and \eqref{51}. The dispersion parameter $b$ has an important effect on the solution \eqref{51}, interestingly it is noticed that the  waves with greater amplitude moves slower than the waves with less amplitude. The obtained solutions are also checked by using Mathematica software and to the best of our knowledge they are not reported earlier in any other research work. The newly obtained solutions may shed light which are governed by KP-BBM equation.

\textbf{\underline{Statements and Declarations}:}\\
\textbf{Acknowledgement:} The authors are grateful to NIT Durgapur, India for their research support.\\
\textbf{Data Availability Statement:} The data that supports the findings of this study are available within the article. \\
\textbf{Conflict of Interest:} The authors declare that they have no conflict of interest.\\
\bibliography{Biblyography}
\bibliographystyle{unsrt}
\end{document}